\documentclass[aps,pra,reprint,longbibliography]{revtex4-1}

\usepackage{graphicx}
\usepackage{amsmath}
\usepackage{amssymb}
\usepackage{textcomp}

\begin{document}
	
\title[]{Superconducting Optoelectronic Neurons \uppercase\expandafter{\romannumeral 2 \relax}: Receiver Circuits}
	
\author{Jeffrey M. Shainline, Sonia M. Buckley, Adam N. McCaughan, Manuel Castellanos-Beltran, Christine A. Donnelly, Michael L. Schneider, Richard P. Mirin, and Sae Woo Nam}
\affiliation{National Institute of Standards and Technology, 325 Broadway, Boulder, CO, 80305}		
	
\date{\today}
	
\begin{abstract}
Circuits using superconducting single-photon detectors and Josephson junctions to perform signal reception, synaptic weighting, and integration are investigated. The circuits convert photon-detection events into flux quanta, the number of which is determined by the synaptic weight. The current from many synaptic connections is inductively coupled to a superconducting loop that implements the neuronal threshold operation. Designs are presented for synapses and neurons that perform integration as well as detect coincidence events for temporal coding. Both excitatory and inhibitory connections are demonstrated. It is shown that a neuron with a single integration loop can receive input from 1000 such synaptic connections, and neurons of similar design could employ many loops for dendritic processing. 
\end{abstract}
	
	
\maketitle

	
\section{\label{sec:introduction}Introduction}
Biological neural networks are sophisticated circuits that receive abundant, temporally varying, disparate information while maintaining complex, internal dynamical states. Such systems can assimilate extraordinary amounts of information over a wide range of time scales, and subsequent exposure to small subsets of information can lead to the successful recall of weak associative memories. Neural computing \cite{scpo2017} aims to capture many of these powerful information processing tools by emulating nature's hardware at the device, circuit, and system levels.

The computational primitives of a neural system are neurons \cite{daab2001,sqbe2008}, relaxation oscillators \cite{st2015,mist1990,soko1993,lued1997,huya2000,bu2006,gile2011,vepe1968,cacl1981} that sum the inputs from many other neurons and, upon reaching a threshold, produce a pulse that is sent to many downstream connections. The concept of using photonic signals with superconducting electronics to form networks of neurons was proposed in Ref.\,\onlinecite{shbu2017}, but that work left many details undeveloped. The principal benefits of using light are the fanout and speed of communication. Superconducting detectors and electronics offer energy efficiency, information processing, and memory. To harness these advantages for neural computation, specific optoelectronic devices must be designed to perform the necessary neural operations.

The goal of this series of papers is to develop the specific superconducting optoelectronic devices that may be used for high-performance neural systems, and the focus of this paper is on the conversion of photonic communication events on many synapses to an integrated total signal stored in the neuron. These optoelectronic devices must meet several criteria: 1) The neuron must be able to achieve leaky integrate-and-fire functionality \cite{daab2001,geki2002} wherein activity on multiple synapses contributes to an integrated signal with a controllable leak rate; 2) Single-photon detection events must contribute to the integrated signal, and the amount each detection event contributes to the integrated signal should depend on a dynamically reconfigurable synaptic weight; 3) Neurons that are sensitive to the sum of spike events must be achievable in order to make use of rate-coded signals \cite{st1967}, and neurons that are sensitive to the timing between afferent spikes must also be achievable in order to make use of temporal coding \cite{thde2001,geki2002,sase2001,stgo2005}; 4) The circuits must scale to thousands of synaptic connections to integrate information across moderately sized cognitive circuits \cite{sh2018e}; 5) The dynamic range of the neuron and synapses should allow activity on a large fraction of the synapses to contribute to a neuronal firing event, yet repeated activity on a small fraction of the synapses should also be able to induce a neuronal firing event; 6) Synapses with inhibitory as well as excitatory functionality must be achievable, and inhibition must work in conjunction with dendritic spines \cite{budr2004,bu2006,haah2015} to enable synchronization on multiple time scales \cite{sase2001,enfr2001,vala2001,budr2004,robu2015}. This paper explores circuit designs satisfying all these criteria.

We design a device that performs the operation of transducing single-photon signals to supercurrent stored in a loop. The circuit utilizes superconducting-nanowire single-photon detectors (SPDs) \cite{gook2001,nata2012,liyo2013,mave2013} in conjunction with Josephson junctions (JJs) \cite{ti1996,vatu1998,ka1999} and mutual inductors \cite{miha2005} to achieve the desired operations. Modification of a current bias can change the synaptic weight of the connection, and we present designs for receiver circuits in neurons with 10, 100, and 1000 synaptic connections. Each synapse is an analog photon-to-fluxon transducer wherein the number of fluxons produced upon the detection of a photon is proportional to the synaptic weight. In Ref.\,\onlinecite{sh2018c} we discuss how these receiver circuits can be coupled to similar SPD/JJ circuits to achieve dynamic synaptic weights capable of spike-timing-dependent plasticity and metaplasticity. In Ref.\,\onlinecite{sh2018d} we discuss how the integrated supercurrent can be compared to a threshold, and the output circuit used to drive a nanophotonic light source. 
	
\section{\label{sec:conceptualOverview}Conceptual overview}
\begin{figure} 
	\centerline{\includegraphics[width=8.6cm]{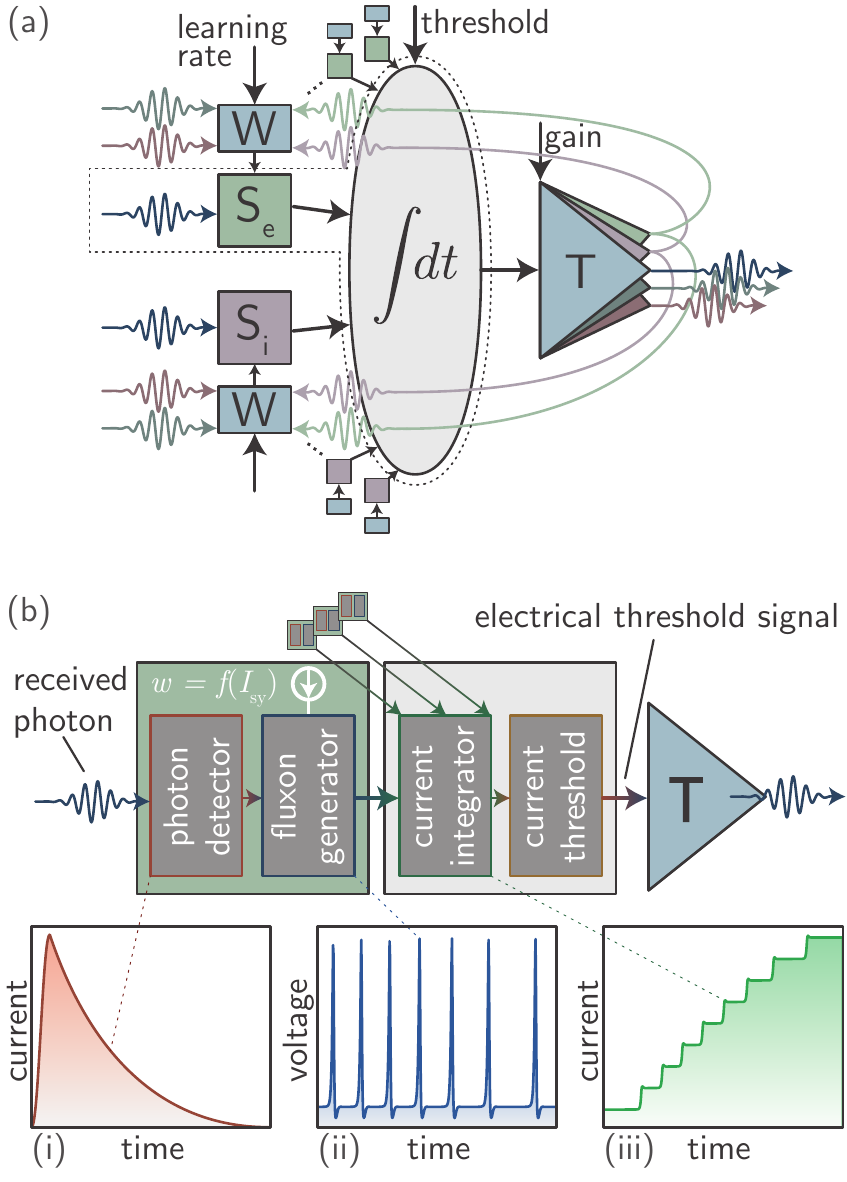}}
	\caption{\label{fig:receivers_schematic}(a) Schematic of the neuron showing excitatory ($\mathsf{S_e}$) and inhibitory synapses ($\mathsf{S_i}$) connected to an integration loop with a variable threshold. The wavy, colored arrows are photons, and the straight, black arrows are electrical signals. The output of the thresholding integrator is input to the transmitter, $\mathsf{T}$. The dashed box encloses a synapse and the integration loop, which are the focus of this work. (b) Sequence of events during synaptic firing event. (i) Single-photon detector transduces photon to electrical current. (ii) Fluxons produced when SPD diverts current to JJ. The number of fluxons is determined by the synaptic bias current, which is controlled by the box labeled $\mathsf{W}$ in (a), discussed in Ref.\,\onlinecite{sh2018c}. (iii) Fluxons added to integration supercurrent storage loop. When the current in the loop reaches threshold, an electrical signal is sent to the transmitter.}
\end{figure}
A schematic of the neuron under consideration is shown in Fig.\,\ref{fig:receivers_schematic}(a). Operation is as follows. Photons from afferent neurons are received by SPDs at a neuron's synapses. Using Josephson circuits, these detection events are converted into an integrated supercurrent that is stored in a superconducting loop. The amount of current added to the integration loop during a synaptic photon detection event is determined by the synaptic weight. The synaptic weight is dynamically adjusted by another circuit combining SPDs and JJs \cite{sh2018c}. When the integrated current from all the synapses of a given neuron reaches a threshold, an amplification cascade begins, resulting in the production of light from a waveguide-integrated light-emitting diode. The photons thus produced fan out through a network of passive dielectric waveguides and arrive at the synaptic terminals of other neurons where the process repeats.

The synaptic receiver circuit, which is the focus of this work, is enclosed in the dashed box of Fig.\,\ref{fig:receivers_schematic}(a). The operation of the synapse within the dashed box is illustrated schematically in Fig.\,\ref{fig:receivers_schematic}(b). The photons produced when a neuron fires are received at downstream synaptic connections by an SPD. At a single synaptic connection, an SPD converts a photon to an electrical signal, namely an electrical supercurrent (Fig.\,\ref{fig:receivers_schematic}(b) part (i)). This supercurrent is diverted from the SPD to a JJ, where it causes the net current through the JJ to exceed $I_c$, temporarily switching the junction to the voltage state and generating a series of fluxons (Fig.\,\ref{fig:receivers_schematic}(b) part (ii)). We refer to this detection of a photon by the SPD and subsequent generation of fluxons by the JJ as a synaptic firing event. The synaptic weight of the connection is implemented via the current bias across the JJ. The effect of this synaptic weight is to change the duration the JJ is held in the voltage state, and therefore the number of fluxons generated during a synaptic firing event. If the synaptic weight is weak, a small number of fluxons, and therefore a small total amount of supercurrent, will be generated during the synaptic firing event and added to the integration loop. If the synaptic weight is strong, a large number of fluxons, and therefore a large amount of supercurrent, will be generated during the synaptic firing event. The means to control this synaptic bias current are discussed in Ref.\,\onlinecite{sh2018c}. The SPD response is virtually identical whether the number of photons present is one or greater than one, and for energy efficiency it is advantageous to send the fewest number of photons possible to each synaptic connection. The SPD response also does not depend strongly on the frequency of light across a bandwidth broad enough for multiplexing \cite{mave2013}. Implementing synaptic weight in the electronic domain in this manner makes use of both the speed and energy efficiency of Josephson junctions, while leveraging the strengths of light for communication.  

The supercurrent generated during each synaptic firing event is added to a superconducting loop, called the synaptic integration (SI) loop, which integrates the total current from all synaptic firing events at that synapse (Fig.\,\ref{fig:receivers_schematic}(b) part (iii)). A neuron may comprise multiple stages of cascaded loops connected by mutual inductors. These neuronal integration (NI) loops combine the signals from all the synapses connected to the neuron. Ultimately, the current coupled to the NI loop(s) is amplified using a current transformer which induces current in a final loop, the neuronal thresholding (NT) loop. The NT loop is a superconducting loop which contains a JJ which produces an output current pulse when its critical current (threshold) is reached. This threshold can be dynamically varied with a current bias. The current pulse generated when the neuron reaches threshold is amplified and ultimately used to trigger a photon\textendash generation event. The amplification stage of the circuit operation is discussed in Ref.\,\onlinecite{sh2018d}.  
	
\section{\label{sec:circuitOperation}Circuit operation}
\begin{figure} 
	\centerline{\includegraphics[width=8.6cm]{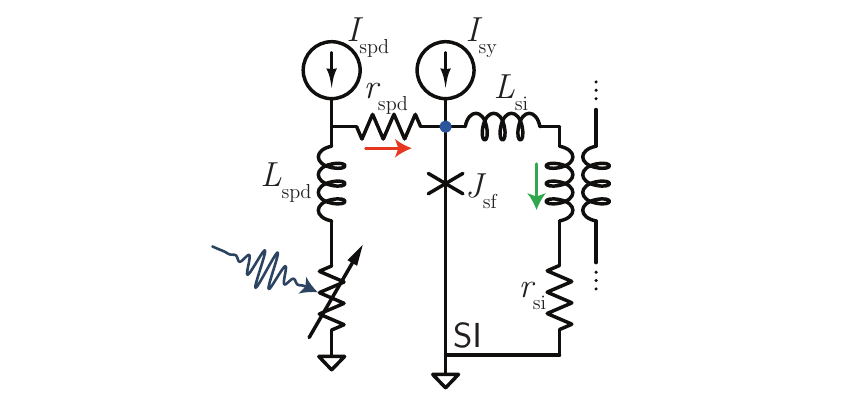}}
	\caption{\label{fig:receivers_simpleCircuit}Circuit diagram of a simple photon-to-fluxon transducer combining a single-photon detector, Josephson junction, and flux storage loop.}
\end{figure}

As described in Sec. \ref{sec:conceptualOverview}, the synaptic receiver circuit under consideration transduces photonic signals to stored supercurrent. A simple instantiation of the concept is shown in Fig.\,\ref{fig:receivers_simpleCircuit}, and a variant without a JJ is discussed in Appendix \ref{apx:sffgNoJJ}. The transduction portion of the synapse comprises an SPD in parallel with a JJ. The SPD is shown as a variable resistor in series with an inductor, and it is current-biased with $I_{\mathrm{spd}}$. The variable resistor has zero resistance in the steady state, and it switches to a high-resistance state temporarily upon absorption of a photon \cite{yake2007}. The synaptic-firing JJ ($J_{\mathrm{sf}}$) is current-biased by the DC current $I_{\mathrm{sy}}$. This synaptic bias current is below the junction critical current, $I_{\mathrm{c}}$. In general, JJs are current biased to bring them to the desired operating point relative to $I_c$ \cite{vatu1998,ka1999}. For the photon/fluxon transducer circuit, the sum of the SPD bias current and the synaptic bias current are chosen to exceed $I_c$. Thus, when the SPD detects a photon, it diverts its bias current to the JJ, temporarily switching $J_{\mathrm{sf}}$ to the voltage state, causing the production of flux quanta. The number of flux quanta generated in a synaptic firing event depends on the relation between $I_c$, $I_{\mathrm{spd}}$, and $I_{\mathrm{sy}}$, as well as the SPD time constant, $L_{\mathrm{spd}}/r_{\mathrm{spd}}$. This flux is trapped in the SI loop. Utilization of a JJ in this circuit is advantageous to decouple the amount of current added to the loop from the time it is stored in the loop (see Appendix \ref{apx:sffgNoJJ}). The SI loop is inductively coupled to the NI loop, which receives input from many synapses. The current in the SI loop decays with the $\tau_{\mathrm{si}} = L_{\mathrm{si}}/r_{\mathrm{si}}$ time constant, which can be chosen over a broad range. By choosing $\tau_{\mathrm{si}}$ to be different for different synapses, one can diversify the temporal information provided to the neuron \cite{stsa2000,abre2004,budr2004,be2007} (see discussion of reset and refraction in Ref.\,\onlinecite{sh2018d}).

\begin{figure} 
	\centerline{\includegraphics[width=8.6cm]{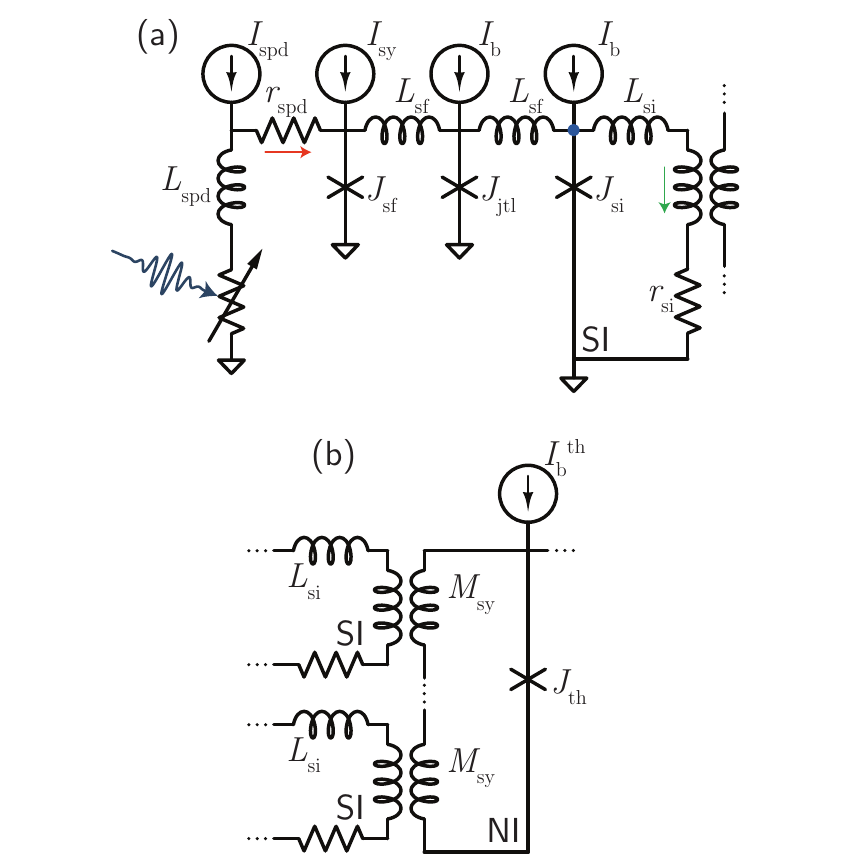}}
	\caption{\label{fig:receivers_circuitDiagrams}(a) Circuit diagram of the photon-to-fluxon transducer connected to the synaptic integration loop by a JTL. (b) Circuit diagram of multiple synapses connected to the neuronal integration loop and the neuronal thresholding loop.}
\end{figure}

The circuit of Fig.\,\ref{fig:receivers_simpleCircuit} captures the concept of the receiver, but its performance is limited in this configuration because the SI loop saturates at a small current. Higher saturation current is achieved by separating the transduction operation from the SI loop by a Josephson transmission line (JTL) \cite{ka1999,vatu1998}, as shown in Fig.\,\ref{fig:receivers_circuitDiagrams}(a). This form of the receiver circuit is the form used as a synapse in this work. 

In the configuration of Fig.\,\ref{fig:receivers_circuitDiagrams}(a), the fluxons produced by the switching of $J_{\mathrm{sf}}$ during a synaptic firing event propagate down a Josephson transmission line (a single JJ in this study), and drive the switching of a junction inside the SI loop.
\begin{figure} 
	\centerline{\includegraphics[width=8.6cm]{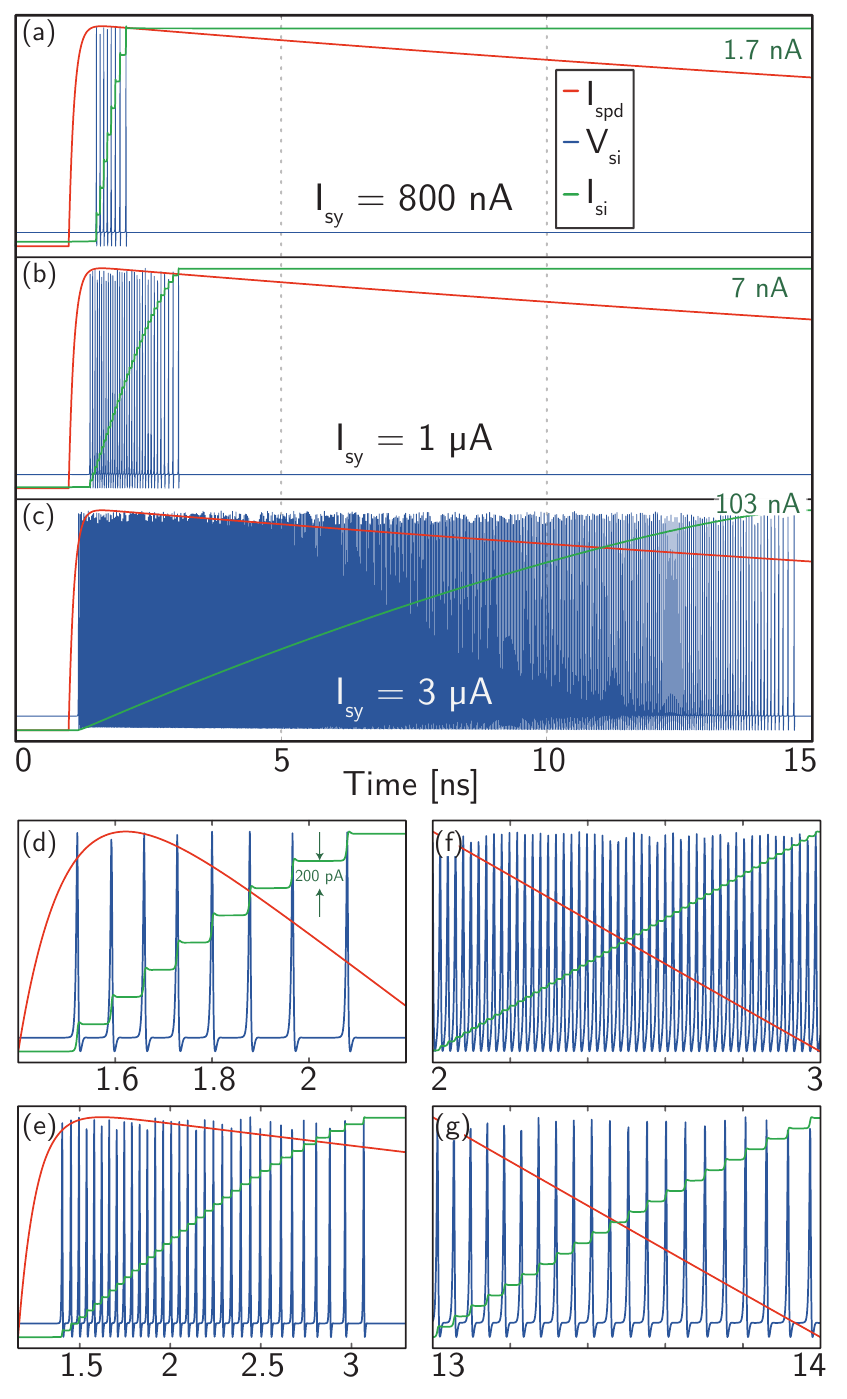}}
	\caption{\label{fig:receivers_sffg_firingDetail} Operation of the synaptic firing circuit during a synaptic firing event for three values of $I_{\mathrm{sy}}$. The three traces in each of these plots are normalized such that the maximum of each trace within the displayed time window is set to one and the minimum is set to zero. The colors of the traces are in reference to the current paths and voltage node labeled in Fig.\,\ref{fig:receivers_circuitDiagrams}(a). (a) Activity of the synaptic firing circuit for $I_{\mathrm{sy}} = 800$\,nA. (b) Activity of the synaptic firing circuit for $I_{\mathrm{sy}} = 1$\,\textmu A. (c) Activity of the synaptic firing circuit for $I_{\mathrm{sy}} = 3$\,\textmu A. (d) Temporal zoom for $I_{\mathrm{sy}} = 800$\,nA. (e) Temporal zoom for $I_{\mathrm{sy}} = 1$\,\textmu A. (f) Temporal zoom near the beginning of the photon-detection pulse for $I_{\mathrm{sy}} = 3$\,\textmu A. (g) Temporal zoom near the end of the photon-detection pulse for $I_{\mathrm{sy}} = 3$\,\textmu A.} 
\end{figure}  
The fluxons from multiple synaptic firing events can be stored in the SI loop, and therefore we may wish to use a loop that can contain many fluxons. The current added to the loop by a single fluxon is $I_{\phi} = \Phi_0/L_{\mathrm{si}}$, where $\Phi_{0} = h/2e = 2.07\times 10^{-15}$\,Wb is a quantum of magnetic flux. The SI loop can maintain a linear response in the presence of many synaptic firing inputs if $L_{\mathrm{si}}$ is chosen to be large, or the SI loop can saturate and act as a high-pass filter if $L_{\mathrm{si}}$ is chosen to be small, thus providing one means of implementing short-term plasticity \cite{abre2004}.

The SI loop is inductively coupled to the NI loop, (Fig.\,\ref{fig:receivers_circuitDiagrams}(b)), which stores a current proportional to the sum of the currents in all the SI loops. The use of mutual inductors allows many synapses to add current to an NI loop without introducing leakage current pathways. Finally, the NI loop couples to a third loop, the NT loop. The mutual inductor coupling the NI loop to the NT loop serves as a transformer to amplify the current, increasing the current that must be detected at threshold. The NT loop may not need to be a separate loop when the number of synapses, $N_{\mathrm{sy}}$, is small. The performance of the NT loop upon reaching the current threshold is discussed in Ref.\,\onlinecite{sh2018d}. A neuron of this variety integrates the combined currents from all of the synapses and communicates that signal to the thresholding loop, as shown in Fig.\,\ref{fig:receivers_circuitDiagrams}(b). 

In Fig.\,\ref{fig:receivers_sffg_firingDetail}, we demonstrate the operation of the synaptic receiver as it experiences a synaptic firing event. Here we use WRSpice \cite{wh1991} to model the circuit of Fig.\,\ref{fig:receivers_circuitDiagrams}(a). We treat the SPD as a current source with exponential rise with 100 ps time constant followed by exponential decay with 50 ns time constant. The amplitude of the SPD current pulse is 10\,\textmu A (see Appendix \ref{sec:appendix_sffg}). Figures \ref{fig:receivers_sffg_firingDetail}(a)-(c) show the activity of a synaptic firing event for $I_{\mathrm{sy}} = 800$\,nA, 1\,\textmu A, and 3\,\textmu A, respectively. With $I_{\mathrm{sy}} = 800$\,nA, the junction is briefly driven above $I_c$, and eight fluxons are transmitted to the SI loop. The synaptic firing event causes the current in the SI loop, $I_{\mathrm{si}}$, to increase by 1.7\,nA. If we increase the bias current by an amount equal to six times the thermal current noise (assuming $L_{\mathrm{spd}} = 100$\,nH), then $I_{\mathrm{sy}} = 1$\,\textmu A. The synaptic firing event produces 33 fluxons and adds 7\,nA to the SI loop. Further increasing the synaptic bias to 3\,\textmu A gives the behavior shown in Fig.\,\ref{fig:receivers_sffg_firingDetail}(c). In this case, 497 fluxons add 103\,nA to the SI loop. 

The period of the voltage pulses is observed to decrease through the duration of the SPD pulse. When the JJ is maintained in the voltage state, a flux quantum will be produced when $\int V(t) dt = \Phi_0$. As the current from the SPD pulse decays, the voltage across $J_{\mathrm{sf}}$ decreases, leading to a longer duration between production of flux quanta. The energy consumed by a synaptic firing event is discussed in Appendix \ref{sec:appendix_energy}.

The analysis of Fig.\,\ref{fig:receivers_sffg_firingDetail} provides the currents and voltages present during a synaptic firing event for three values of $I_{\mathrm{sy}}$. Appendix \ref{sec:appendix_sffg} quantifies the device performance more systematically. A principal objective of this analysis is to determine the range of synaptic bias currents over which we would like to operate. Operating with a minimum synaptic bias of 1\,\textmu A enables us to work close to the energy-efficiency limit of the circuit, and we anticipate that the exact number of fluxons produced during a firing event will be noisy, much like the activity of a biological neuron \cite{stgo2005}. The amount of current added to the SI loop during a synaptic firing event with strong synaptic weight should be significantly larger than the amount of current with a weak synaptic weight. We choose $I_{\mathrm{sy}} = 3$\,\textmu A to be the largest synaptic bias at which we would like to operate, and thus a synaptic firing event with a strong synaptic bias adds 15 times as much current to the SI loop (and therefore the NI loop and NT loop) as a firing event with a weak synaptic bias. This ratio is entirely tunable based on the needs of the system. Learning\textemdash either supervised or unsupervised \textemdash should adjust the synaptic bias current over the range 1\,\textmu A $< I_{\mathrm{sy}} < 3$\,\textmu A. Circuits accomplishing this are discussed in Ref.\,\onlinecite{sh2018c}.

\section{\label{sec:scaling}Multisynaptic neurons}
In general, a neuron will combine signals from many synaptic connections and produce a pulse when this combined signal reaches a threshold. We would like to know how devices will perform when many synapses are integrated with a single NI loop. Combining the data from Figs. \ref{fig:receivers_sffg_firingDetail} and \ref{fig:receivers_IsilPerPhoton} (Appendix \ref{sec:appendix_sffg}), we find that the SI loop with this design can receive over 1000 synaptic firing events when $I_{\mathrm{sy}} = 1$\,\textmu A, and 82 synaptic firing events when $I_{\mathrm{sy}} = 3$\,\textmu A before saturation of the loop occurs (assuming $\tau_{\mathrm{si}}\rightarrow \infty$). If the loop contains a resistance, the trapped flux will leak with the $L/r$ time constant, leaving the synapse ready to receive further synaptic firing events. We wish to determine whether or not this activity is sufficient to produce a neuronal firing event. To conduct this analysis, we need to analyze how the SI loops couple to the NI loop and the NT loop. These calculations are described in Appendix \ref{sec:appendix_ntl}.  

\begin{figure} 
	\centerline{\includegraphics[width=8.6cm]{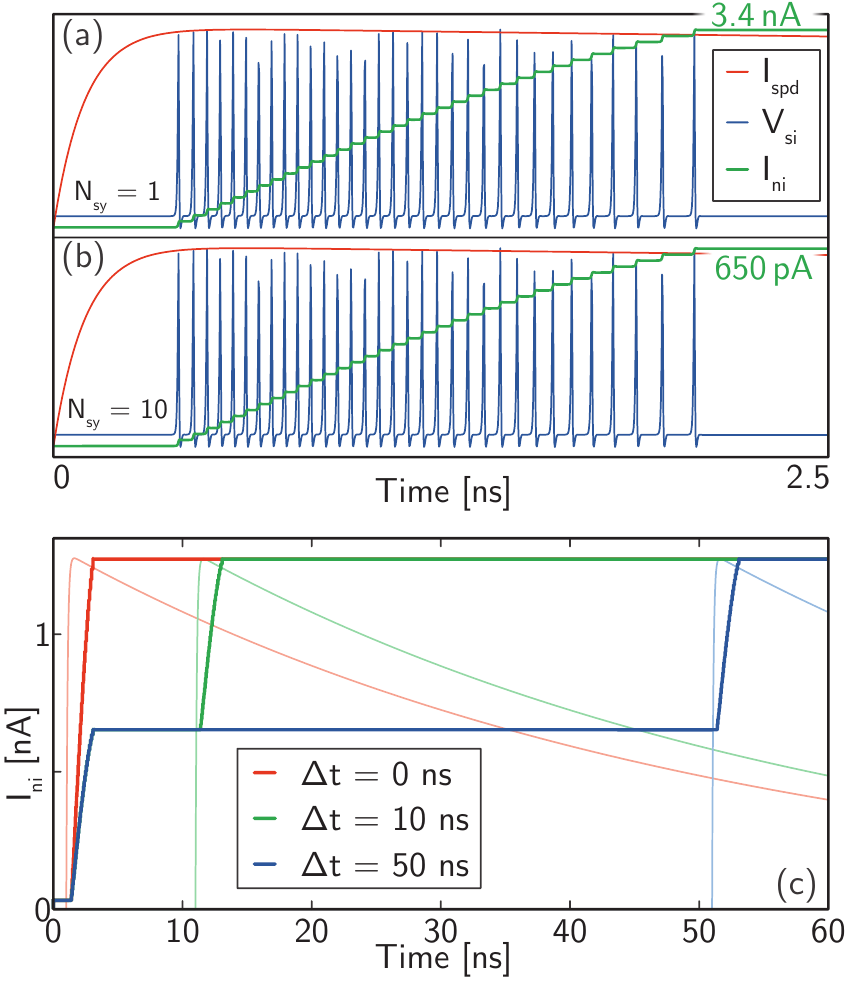}}
	\caption{\label{fig:receivers_Nsy10}Variation of neuronal response with number of synaptic connections. (a) Activity during a synaptic firing event when only a single synapse is present. (b) Activity during a synaptic firing event when ten excitatory synapses are present. (c) Effect of timing delay between two synaptic firing events on different synapses in a neuron of $N_{\mathrm{sy}} = 10$. The red traces show the SPD pulses and current added to the neuronal integration loop for two coincident synaptic firing events. The green traces show a delay of 10 ns between the two synaptic firing events. The blue traces show a delay of 50 ns between the two synaptic firing events. In all three scenarios, the total accumulated current in the neuronal integration loop is equal.}
\end{figure}
In Appendix \ref{sec:appendix_ntl} we motivate choices for $L_{\mathrm{si}}$ and $M_{\mathrm{sy}}$ (Fig.\,\ref{fig:receivers_circuitDiagrams}(b)). We wish to know how inductively coupling multiple SI loops to a single NI loop affects the operation during synaptic firing events. Figures \ref{fig:receivers_Nsy10}(a) and (b) compare a synaptic firing event of a neuron with a single synaptic connection to a synaptic firing event of a single synapse connected to an NI loop with 10 synaptic connections. The number and timing of the fluxons is identical. The effective inductance of the SI loops in the two cases is dominated by $L_{\mathrm{si}}$, so the current added to the SI loop with each fluxon is nearly independent of the number of synaptic connections on the NI loop. The amount of current added to the NI loop with each fluxon added to a SI loop depends on the ratio of $M_{\mathrm{sy}}$ to the total inductance of the NI loop, which depends on $N_{\mathrm{sy}}$. After a single synaptic firing event on a NI loop with $N_{\mathrm{sy}} = 1$, $I_{\mathrm{ni}} = $ 3.4\,nA, and after a single synaptic firing event on a NI loop with $N_{\mathrm{sy}} = 10$, $I_{\mathrm{ni}} = $ 650\,pA. These examples are for the case of $I_{\mathrm{sy}} = 1$\,\textmu A, but the conclusions hold across the operational range of $I_{\mathrm{sy}}$. The purpose of Fig.\,\ref{fig:receivers_Nsy10}(a) and (b) is to demonstrate that SI loops inductively coupled to an NI loop do not change their behavior when additional synapses are added to the neuronal loop. 

The effect of timing delay between two synaptic firing events on different synapses in a neuron of $N_{\mathrm{sy}} = 10$ is shown in Fig.\,\ref{fig:receivers_Nsy10}(c). The total current added to the NI loop is independent of the timing delay between the two synaptic firing events. These linearities with respect to $N_{\mathrm{sy}}$ and pulse timing delay are attractive features of inductively coupled synapses. Contexts in which nonlinearity with respect to arrival time is desirable, such as for temporal coding \cite{thde2001,sase2001} or dendritic processing \cite{brcl2010,haah2015}, are likely to employ two-photon receiver circuits such as discussed in Fig.\,\ref{fig:receivers_coincidence}. The purpose of Fig.\,\ref{fig:receivers_Nsy10}(c) is to demonstrate that SI loops inductively coupled to an NI loop do not change their behavior when additional synapses on the neuronal loop fire concurrently. The independent behavior of the SI loops when inductively coupled to the NI loop are an important reason why inductive coupling is preferable to direct wiring of synapses to a common firing JJ or integration element.
 
\begin{figure} 
	\centerline{\includegraphics[width=8.6cm]{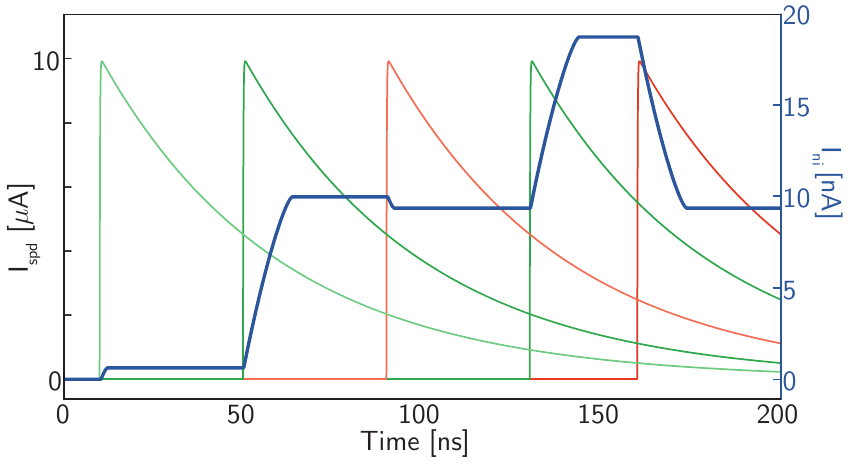}}
	\caption{\label{fig:receivers_inhibitoryAndExcitatory}A neuron with seven excitatory and three inhibitory synaptic connections. The excitatory and inhibitory current inputs are shown as green and red traces and are referenced to the left $y$ axis. The blue trace is $I_{\mathrm{sy}}$, referenced to the right $y$ axis. At time $t = 10$ ns, a synaptic firing event occurs on an excitatory synapse with $I_{\mathrm{sy}} = 1$\,\textmu A. At time $t = 50$\,ns, a synaptic firing event occurs on an excitatory synapse with $I_{\mathrm{sy}} = 3$\,\textmu A. At time $t = 90$\,ns, a synaptic firing event occurs on an inhibitory synapse with $I_{\mathrm{sy}} = 1$\,\textmu A. At time $t = 130$\,ns, a synaptic firing event occurs on an excitatory synapse with $I_{\mathrm{sy}} = 3$\,\textmu A. At time $t = 160$\,ns, a synaptic firing event occurs on an inhibitory synapse with $I_{\mathrm{sy}} = 3$\,\textmu A. The colors in this plot are not in reference to Figs.\,\ref{fig:receivers_circuitDiagrams} - \ref{fig:receivers_Nsy10}.}
\end{figure}
It is important for a neuron to be able to receive excitatory and inhibitory connections \cite{vrso1996,sase2000,daab2001,robu2015}. Inhibitory connections keep the network from experiencing runaway activity and are crucial for temporal synchronization \cite{sase2001,vala2001,enfr2001,budr2004,bu2006}. Inhibitory connections can be constructed with the same photon-to-fluxon transduction circuit presented thus far by changing the sign of $M_{\mathrm{sy}}$. We investigate a neuron with seven excitatory and three inhibitory connections in Fig.\,\ref{fig:receivers_inhibitoryAndExcitatory}. The figure shows a time trace of $I_{\mathrm{ni}}$ as three excitatory and two inhibitory synaptic firing events occur. One of the excitatory events and one of the inhibitory events occur in synapses with $I_{\mathrm{sy}} = 1$\,\textmu A, and the other events occur in synapses with $I_{\mathrm{sy}} = 3$\,\textmu A. This plot demonstrates the dynamic state of a multi-synaptic neuron under the influence of excitatory and inhibitory connections. 

The symmetry between inhibitory and excitatory synapses is broken by $I_{\mathrm{b}}^{\mathrm{th}}$, the current bias across the thresholding junction. The circuit can be designed so that saturation of all inhibitory SI loops is insufficient to add enough counter current to the NT loop to overcome  $I_{\mathrm{b}}^{\mathrm{th}}$ and reach threshold. Thus, repeated excitatory events can drive the neuron to spike, but repeated inhibitory events can only move the device further from threshold and cannot trigger a spike, much like the polarizing effects of inhibitory interneurons in biological neural systems.
	
\section{\label{sec:dendriticProcessing}Dendritic processing}
In addition to neurons that integrate single-photon pulses, as described in Sec. \ref{sec:circuitOperation}, it is desirable to achieve neurons that detect coincident signals from two or more pre-synaptic neurons for detecting temporally coded information \cite{thde2001,sase2001,stse2007,sp2008,brcl2010,haah2015}. The mutual information regarding a stimulus conveyed by two or more neurons can be approximated by a power series (or Volterra expansion \cite{geki2002}) with the leading term corresponding to firing rate, and the second-order term representing correlations \cite{pasc1999}. In biological neurons, temporal synaptic sequences can be detected using hardware nonlinearities present in dendritic spines \cite{brcl2010,haah2015}, which perform important cortical computations. Detection of timing correlations and sequences can be achieved in the optoelectronic hardware platform under consideration using two (or more) SPDs in a similar circuit to the synaptic receiver of Fig.\,\ref{fig:receivers_circuitDiagrams}(a). 

\begin{figure} 
	\centerline{\includegraphics[width=8.6cm]{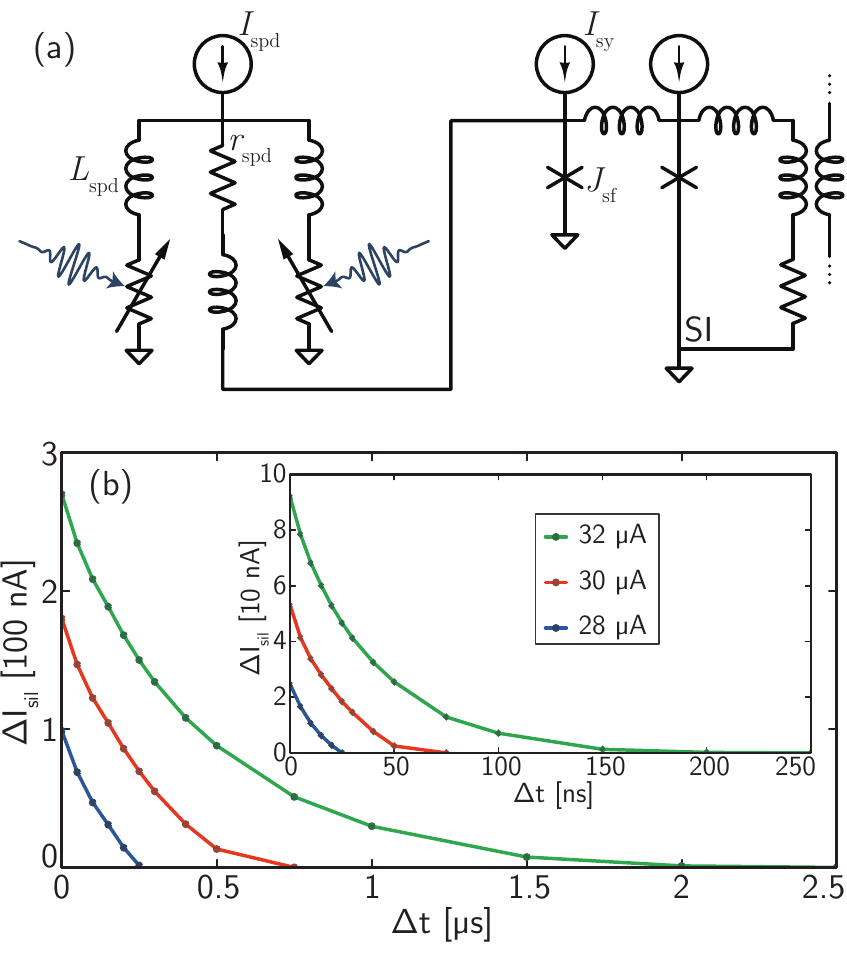}}
	\caption{\label{fig:receivers_coincidence}Coincidence-detection circuit for neurons sensitive to temporal coding. (a) Circuit diagram of symmetric two-photon coincidence detection circuit. (b) Current added to SI loop as a function of time delay between arrival of the photons. Performance was calculated for the three values of $I_{\mathrm{sy}}$ shown in the legend.}
\end{figure}

In Fig.\,\ref{fig:receivers_coincidence} we analyze a two-photon symmetrical coincidence detection circuit. The circuit diagram is shown in Fig.\,\ref{fig:receivers_coincidence}(a). The two SPDs are biased symmetrically, and the circuit is designed such that if either SPD detects a photon in isolation, the current across the JJ remains below $I_c$, but if both detect a photon within a certain time window, the current across the JJ can exceed $I_c$, adding current to the SI loop. The amount of current added to the SI loop is plotted as a function of the difference in arrival times between two photons in Fig.\,\ref{fig:receivers_coincidence}(b). WRSpice was again used for these simulations, but in this case the SPDs were modeled not as current sources, but as resistors of 5\,k$\Omega$ with 200\,ps duration occurring at specified photon-arrival times \cite{yake2007}. The time scale over which correlated events are detected is set by the $L_{\mathrm{spd}}/r_{\mathrm{spd}}$ time constant of the circuit. In the main panel, this time constant is 500\,ns, and in the inset it is 50\,ns. Longer correlation windows can be straightforwardly achieved, and the shortest correlation window will be limited by the latching time of the SPD.

\begin{figure} 
	\centerline{\includegraphics[width=8.6cm]{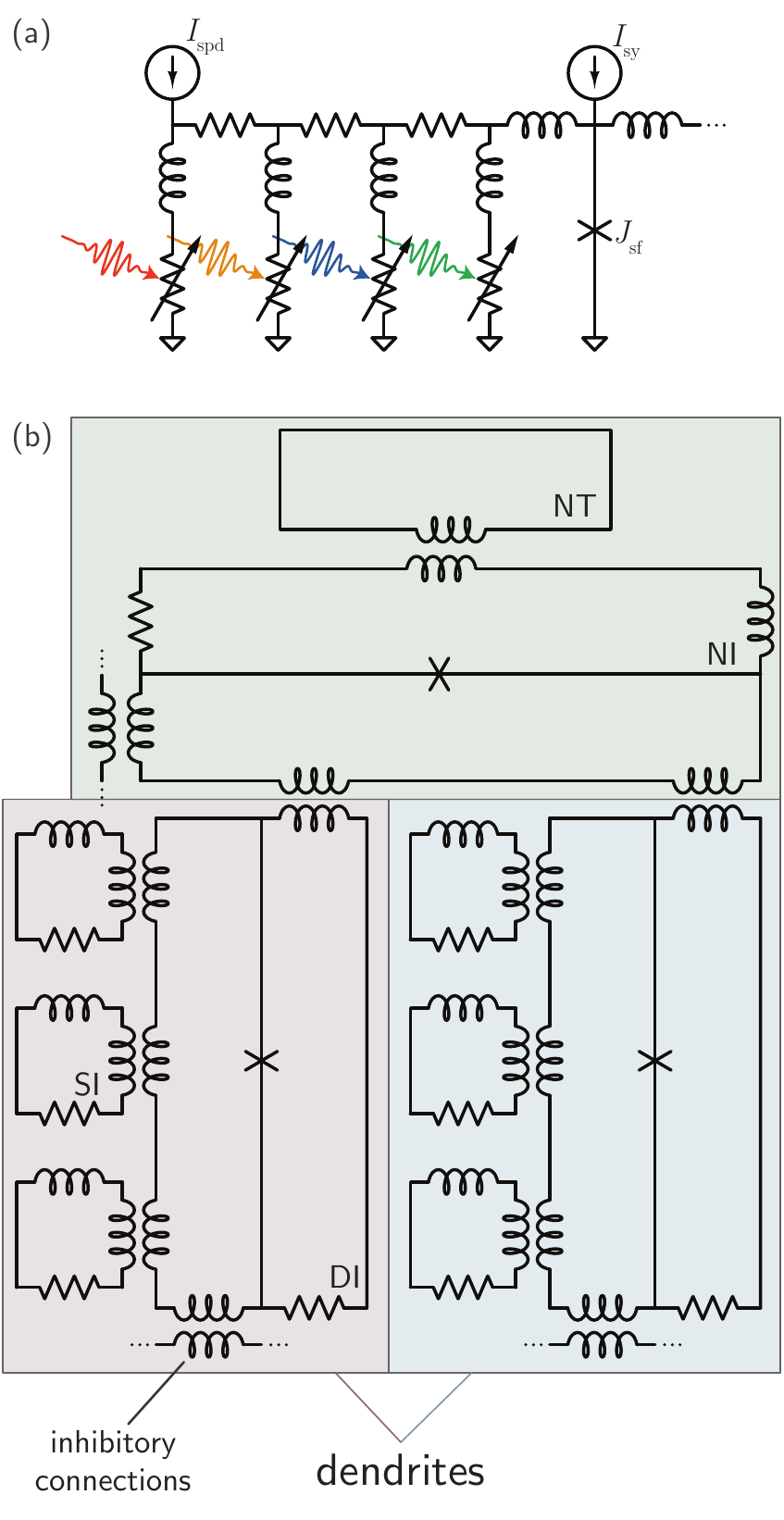}}
	\caption{\label{fig:receivers_dendriticProcessing} Two approaches to dendritic processing. (a) Schematic of optoelectronic circuit for detecting a specific sequence of synaptic events. (b) Schematic of hierarchy of loops for nonlinear electrical response. In practice, JJs in the DI loops would need to be current biased, and a buffer stage would likely be inserted into loops contacted by inhibitory connections.}
\end{figure}

Due to the symmetric biasing of the two SPDs, the circuit of Fig.\,\ref{fig:receivers_coincidence} is insensitive to order of photon arrival. By breaking this symmetry, similar receiver circuits which detect ordered correlations can be used for Hebbian learning \cite{sh2018c}. The two-SPD circuit of Fig.\,\ref{fig:receivers_coincidence} can also be extended to detect other sequences of activity, as shown in Fig.\,\ref{fig:receivers_dendriticProcessing}(a), wherein a specific sequence of four photons will trigger synaptic firing. In this example of using optoelectronic circuits to emulate nonlinear processing in dendrites, the cascade of photon detectors plays a role analogous to a dendritic spine, which may bring the neuron closer to threshold if a specific sequence of photons is incident on the SPD array. In this case, only the sequence red\textendash yellow\textendash blue\textendash green will add flux to the SI loop. Each SPD plays the role of a synapse, and the SPDs taken together with $J_{\mathrm{sf}}$ and the SI loop play the role of a dendritic spine. 

Dendritic processing can also be used for intermediate nonlinear processing between synapses and the neuron \cite{sava2017}. An example circuit is shown in Fig.\,\ref{fig:receivers_dendriticProcessing}(b). Here, multiple SI loops are inductively coupled to another loop, which contains a JJ. Only when the junction is driven above its critical current does an appreciable signal get added to the dendritic integration (DI) loop, which is an intermediate, nonlinear processor between the SI loops and the NI loop. In this case, the DI loops are analogous to dendritic spines. An important role of dendritic spines is in conjunction with inhibitory interneurons that can temporarily suppress the efficacy of an entire dendritic spine \cite{budr2004,bu2006,robu2015}. At the bottom of Fig.\,\ref{fig:receivers_dendriticProcessing}(b) we show how an inhibitory interneuron may be inductively coupled to a dendritic spine. When inhibition is applied to the loop, it may be impossible for the synaptic connections to drive the JJ above threshold and add flux to the DI loop. Many levels of loop hierarchies can be combined in this way to achieve various nonlinear functions as well as current amplification before the neuronal thresholding loop.

As discussed in Ref.\,\onlinecite{sh2018a}, dendritic processing in conjunction with inhibitory interneurons contributes to network synchronization on various temporal and spatial scales \cite{vala2001,enfr2001,sase2001,bu2006,buwa2012,robu2015}. The approach to dendritic spines shown in Fig.\,\ref{fig:receivers_dendriticProcessing}(b) is one way inhibition could be used with the synapses presented here to achieve these functions thought to be necessary for cognition \cite{budr2004,fr2015}. In this context, engineering synaptic and dentritic circuits with a variety of time constants (analogous to membrane time constants) is important, as these time constants affect synchronization frequency \cite{lued1997} and enable neurons with a greater diversity of synapses \cite{ma2016}. As discussed in Ref.\,\onlinecite{sh2018a}, power-law dynamics are necessary for information integration and self-organized criticality, and a power-law frequency distribution can be achieved through the superposition of exponential decay functions with a diversity of time constants \cite{be2007}. To achieve this with the dendritic processors shown in Fig.\,\ref{fig:receivers_dendriticProcessing}(b), resistors are placed in each DI loop. The $L/r$ time constant of each DI loop will set its temporal response, and in this way, different dendrites can be given different time constants. Similarly, a resistor can be placed in each SI loop so that each synaptic excitation has a characteristic time constant, as discussed previously. These resistors will also accomplish the task of purging flux from the SI and DI loops to avoid saturation. As indicated in Fig.\,\ref{fig:receivers_dendriticProcessing}(b), inhibition can be applied at various points in the loop hierarchy, including specific synaptic loops, dendritic loops, the neuronal integration loop, and even the current source to the light emitter \cite{sh2018d}. These different structural implementations of inhibition are analogous to the three main forms of inhibition observed in biological neurons, wherein interneurons target dendrites, the soma, and the axon initial segment \cite{robu2015}. 
	
\section{\label{sec:discussion}Discussion}
The present work has investigated a superconducting optoelectronic neuron receiver circuit utilizing an analog SPD-to-JJ transducer that couples flux to a storage loop. The synaptic weight can be enacted by changing the bias to the JJ, and the storage loop can hold flux resulting from between 80 and 1000 synaptic firing events depending on the synaptic weight. It has further been shown that 1000 of these synapses can be inductively coupled to an integration loop and ultimately to a thresholding JJ. Designs for single-photon-sensitive receivers capable of operating on rate-coded signals as well as two-photon-sensitive receivers capable of operating on temporally coded signals have been discussed. Excitatory as well as inhibitory behavior has been demonstrated, and a hierarchy of loops for dendritic processing has been shown.

Variations on the designs presented here may also be useful. For example, we have been discussing varying $I_{\mathrm{sy}}$ to update the synaptic weight, but it is possible to achieve similar functionality by varying $I_{\mathrm{spd}}$ while keeping $I_{\mathrm{sy}}$ fixed. Variations trading dynamic range and/or noise for improved energy efficiency are also possible.

Many aspects of this circuit require further analysis. The task of adjusting the synaptic weight via the synaptic bias current is addressed in Ref.\,\onlinecite{sh2018c}.  A means by which the thresholding JJ can trigger a neuronal firing event is treated in Ref.\,\onlinecite{sh2018d}. A network of waveguides capable of connecting neurons with thousands of synapses is developed in Ref.\,\onlinecite{sh2018e}. That reference also analyzes the spatial scaling of the photonic and electronic devices involved. Throughout these papers, we have emphasized the advantages of short communication delays achieved through light-speed signaling. Yet delays in cortical circuits can be leveraged for computation. One role played by delays is to diversify the oscillation frequencies of various neuronal assemblies in which a given neuron participates. In this context, it is not simply the communication delay due to axon conduction that is responsible for establishing the time constant, but rather the entire delay between a neuronal firing event and the subsequent synaptic firing event at the post-synaptic neuron. In the synaptic receiver circuits presented here, synaptic delays may be introduced in the electronic domain with $LC$ circuits between the synaptic firing junction, $J_{\mathrm{sf}}$, and the SI loop (see Fig.\,\ref{fig:receivers_circuitDiagrams}(a)). This approach will enable networks to diversify oscillation frequencies while maintaining the advantages of short communication delays for high-frequency synchronization \cite{budr2004,bu2006,fr2015} and large-scale information integration \cite{stsa2000}.

In addition to utilization as a neural computer, experiments using these neural circuits may be useful for testing hypotheses in neuroscience. The circuits presented here can be reconfigured and extended to make use of numerous SPDs performing nonlinear correlation functions on signals from numerous pre-synaptic neurons as well as employing multiple integration loops, multiple thresholding elements, and multiple light sources suitable for experimenting with different synaptic and dendritic circuit paradigms. Additionally, similar receiver circuits can be utilized for dynamic learning, synaptic plasticity, and metaplasticity, as presented in Ref.\,\onlinecite{sh2018c}.

In Ref.\,\onlinecite{sh2018a} we argue that cognitive systems benefit from information integration across spatial and temporal scales. Temporal integration is achieved with a power law distribution of neural oscillation frequencies. The receiver circuits presented in this work enable this functionality in at least two ways. First, they are fast and can detect photon communication events at 20 MHz and possibly faster. The brain oscillates at frequencies from 0.05\,Hz to 600\,Hz \cite{budr2004}. Noise is likely to limit superconducting optoelectronic device operation at low frequencies, but we assume these circuits can oscillate down to 1\,Hz. Thus, while the human brain oscillates at frequencies spanning four orders of magnitude, these receivers could contribute to oscillations across seven orders of magnitude, indicating the potential for information integration across very large networks \cite{stsa2000}. The second manner in which these receivers are well-suited to achieving a power law frequency distribution is that their oscillatory response is tunable, so each neuron can participate in a broad range of oscillations. This tunable response is achievable by changing the threshold of the JJ in the NI loop or DI loops by changing bias currents, and also by changing which synapses are effective at a given time using inhibition and dendritic processing. Such dynamic effects in synapses and neurons in the brain are crucial for maximally utilizing the time domain for information integration \cite{bu2006}.

Finally, we point out that while the circuits presented here utilize photons for communication and to trigger synaptic firing events, similar functionality is achievable using only fluxons. The SPD in Fig.\,\ref{fig:receivers_simpleCircuit} can be replaced with an nTron \cite{mcbe2014}, the gate of which can be driven normal by one or more fluxons \cite{sh2018d}. The same techniques of utilizing a hierarchy of integration loops, dendritic processing, and synaptic weighting can be used in those circuits as well. Achieving the communication necessary for large networks \cite{sh2018e} will be cumbersome with purely electronic circuits. Yet such neurons may fire at rates beyond 10\,GHz with very low power consumption when driving up to $\approx$\,20 synaptic connections. Networks combining electronic and optoelectronic neurons extend the power law degree distribution to lower degree and the power law frequency distribution to higher frequency. While other purely electronic, JJ-based neurons and synapses have been proposed \cite{hias2007,crsc2010,ru2016} and demonstrated \cite{segu2014,scdo2018}, we point out how the circuits presented here can be converted to purely electrical neurons to illustrate the continuity of electronic and photonic implementations, and to show that networks with both electrical and optical neurons working in conjunction based on the same neural principles and fabrication process can be achieved.

\vspace{0.5em}
This is a contribution of NIST, an agency of the US government, not subject to copyright.
	
\newpage
\appendix

\section{\label{apx:sffgNoJJ}Synaptic transducer without Josephson junction}
\begin{figure} 
	\centerline{\includegraphics[width=8.6cm]{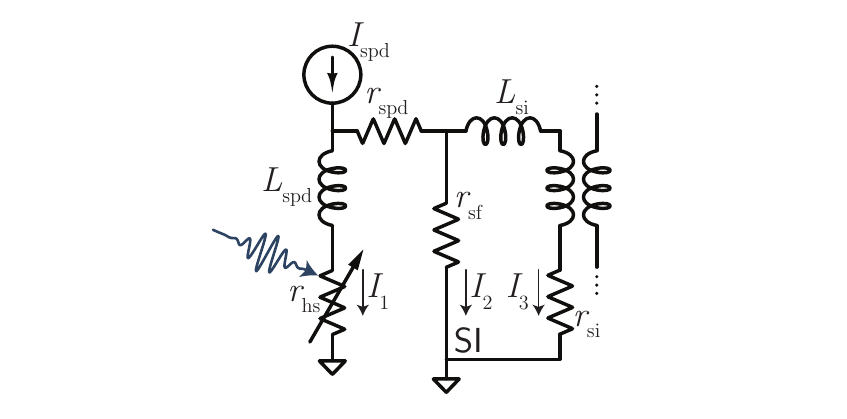}}
	\caption{\label{fig:receivers_sffgNoJJ}Circuit diagram of a simple photon-to-fluxon transducer combining a single-photon detector, resistor, and flux storage loop.}
\end{figure}

A circuit similar to the one shown in Fig.\,\ref{fig:receivers_simpleCircuit} is depicted in Fig.\,\ref{fig:receivers_sffgNoJJ}. 
The equations of motion for this system can be represented as
\begin{equation}
\label{eq:sffgNoJJ_1}
\begin{split}
\frac{dI_1}{dt} & = \frac{1}{L_{\mathrm{spd}}}[ (r_{\mathrm{sf}}+r_{\mathrm{spd}})I_{\mathrm{spd}} \\
& - (r_{\mathrm{sf}}+r_{\mathrm{spd}}+r_{\mathrm{hs}})I_{\mathrm{1}} - r_{\mathrm{sf}}I_{\mathrm{3}}],
\end{split}
\end{equation}
and
\begin{equation}
\label{eq:sffgNoJJ_2}
\frac{dI_3}{dt} = \frac{r_{\mathrm{sf}}}{L_{\mathrm{sil}}}\left( I_{\mathrm{spd}} - I_{\mathrm{1}} - I_{\mathrm{3}} \right).
\end{equation}

The advantage of the circuit in Fig.\,\ref{fig:receivers_sffgNoJJ} is that it can be constructed without Josephson junctions, thereby simplifying the fabrication process. However, this circuit has two primary weaknesses. First, the synaptic bias current, $I_{\mathrm{sy}}$, and the SPD bias current, $I_{\mathrm{spd}}$, cannot be separated, as they could in the case of Fig.\,\ref{fig:receivers_simpleCircuit} because the resistor $r_{\mathrm{sf}}$ disrupts the superconducting path to ground. The synaptic bias must either be limited to the range of the plateau of the SPD \cite{mave2013}, or the synaptic bias will be convoluted with the probability of detecting a photon. Second, the rise time and the decay time of the current $I_{\mathrm{3}}$ are identical (as seen in Eq. \ref{eq:sffgNoJJ_2}), and therefore the amount of current that is added to the SI loop during a synaptic firing event and the lifetime of that current in the SI loop are interdependent. The circuits of Figs. \ref{fig:receivers_simpleCircuit} and \ref{fig:receivers_circuitDiagrams}(a) allow the amount of current that is added to the SI loop (and therefore the synaptic efficacy) to be adjusted in hardware through $L_{\mathrm{sil}}$ and dynamically through $I_{\mathrm{sy}}$, while the photon detection probability is set independently (and dynamically) with $I_{\mathrm{spd}}$, and the synaptic decay time is set independently with $r_{\mathrm{si}}$. 

While it may be useful in the near term to pursue circuits like the one shown in Fig.\,\ref{fig:receivers_sffgNoJJ}, the synaptic weighting circuits described in Ref.\,\onlinecite{sh2018c} cannot be modified in a similar manner to replace Josephson junctions with resistors, as they must store flux indefinitely to maintain memory, and therefore must not have an $L/r$ leak rate. We therefore expect the mature hardware platform to employ Josephson junctions for optimal device performance.

\section{\label{sec:appendix_sffg}Design of synaptic transducer}
Unless otherwise specified, we take $I_{\mathrm{spd}} = 10$\,\textmu A, comparable to the switching current of MoSi \cite{veko2015} SPDs. Designs with lower $I_{\mathrm{spd}}$, as would be present in WSi nanowires \cite{mave2013}, or higher $I_{\mathrm{spd}}$, as would be present in NbN \cite{gook2001} or NbTiN \cite{miya2013} nanowires are also straightforward to achieve. The variable resistor of the SPD has zero resistance in the steady state, and it switches to a high-resistance state ($\approx$\,5\,k$\Omega$) temporarily ($\approx$\,200\,ps) upon absorption of a photon \cite{yake2007}. Typical values for the parameters in Fig.\,\ref{fig:receivers_circuitDiagrams}(a) are $L_{\mathrm{spd}} = 100$\,nH, $I_{\mathrm{spd}} = 10$\,\textmu A, $r_{\mathrm{spd}} = 2$\,$\Omega$, $I_{\mathrm{sy}} = 800$\,nA -  4\,\textmu A, $L_{\mathrm{sf}} = 200$\,pH, $I_{\mathrm{b}} = 7$\,\textmu A - 9\,\textmu A, $L_{\mathrm{si}} = 100$\,nH - 10\,\textmu H, and $M_{\mathrm{sy}} = 1$\,nH. The $I_c$ of the JJs is chosen to be 10\,\textmu A in this work to improve energy efficiency. In Ref.\,\cite{sh2018e} we argue this is not necessary, and implementation with junctions of $I_c = 40$\,\textmu A or higher is probably a better design choice. The JJs used in simulations in this work and the companion paper \cite{sh2018c} have $\beta_c = 0.95$, where $\beta_c = 2eI_cCR^2/\hbar$, with $C$ the junction capacitance and $R$ the junction resistance in the RCSJ model \cite{vatu1998,ka1999}. The parameter $\beta_c$ corresponds to the junction damping (with $\beta_c = 1$ corresponding to critical damping), and for this study, we consider slightly over-damped junctions. Typical values for the amplifying transformer inductors in Fig.\,\ref{fig:receivers_circuitDiagrams}(b) are $L_{\mathrm{at1}} = 1$\,\textmu H and $L_{\mathrm{at2}} = 100$\,pH.

In all flux-storage loops, there is a trade-off between inductance and area. High-kinetic-inductance materials such as WSi have inductance per square as large as 250\,pH/$\square$. By patterning a nanowire of such a material in a meander geometry, we can produce an inductor with 10\,\textmu H in an area of 35\,\textmu m $\times$ 35\,\textmu m with a minimum feature size of 50\,nm. We demonstrate in Ref.\,\onlinecite{sh2018e} that these relatively large inductors are still compatible with scaling to neurons with 1000 synapses because the area of photonic routing is generally the limiting factor.

A JJ coupled to an inductive loop is often characterized by the parameter $\beta_L = 2\pi L I_c/\Phi_0$, which quantifies the amount of phase a loop can store. $\beta_L / 2\pi$ quantifies the number of flux quanta that can be stored. For the design discussed here, $\beta_L/2\pi = 5\times10^4$. For digital computing applications, $\beta_L/2\pi = 1.6$ is typical. There is also an area/inductance trade-off for the mutual inductor coupling each SI loop to the NI loop, and in the present work we choose 1\,nH for this mutual inductor, as will be discussed in Appendix \ref{sec:appendix_ntl}.

Figure \ref{fig:receivers_IsilPerPhoton}(a) and (b) summarize the current added to the SI loop and the number of flux quanta generated during a synaptic firing event for a range of values of $I_{\mathrm {si}}$ and for four values of $L_{\mathrm{sf}}$.
\begin{figure} 
	\centerline{\includegraphics[width=8.6cm]{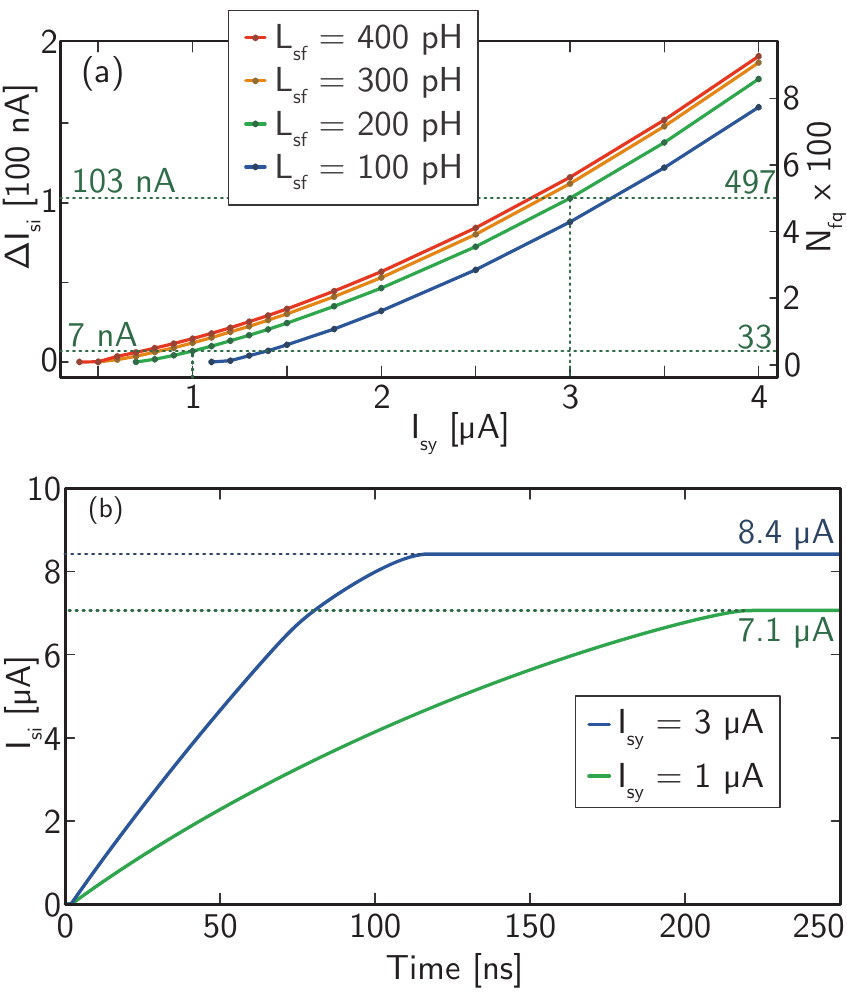}}
	\caption{\label{fig:receivers_IsilPerPhoton} Response of the synaptic integrating loop to photonic activity. (a) Current added to the synaptic integrating loop, $\Delta I_{\mathrm{si}}$, per synaptic firing event as a function of synaptic bias current, $I_{\mathrm{sy}}$. The right $y$-axis gives the corresponding number of flux quanta. We choose to operate with $I_{\mathrm{sy}}$ between 1\,\textmu A and 3\,\textmu A, and the values of $\Delta I_{\mathrm{si}}$ for these operation points are shown on the plot. (b) Current circulating in the synaptic integrating loop as a function of time as flux is added to the loop by fixing the bias of $J_{\mathrm{sf}}$ at 10\,\textmu A $+ I_{\mathrm{sy}}$ for $I_{\mathrm{sy}} = 1$\,\textmu A and 3\,\textmu A. In these calculations, $L_{\mathrm{si}} = 10$\,\textmu H, $I_{\mathrm{spd}} = 10$\,\textmu A, $I_{\mathrm{b}} = 9$\,\textmu A.}
\end{figure}
It is apparent that the nonlinearity of the operation as a function of $I_{\mathrm{sy}}$ is not extreme, indicating that a comfortable range of operating currents can be tolerated, and sensitivity to noise will be not be cumbersome. The dashed green lines indicate reasonable choices for the minimum and maximum values of $I_{\mathrm{sy}}$. At the minimum value of 1\,\textmu A, roughly 30 fluxons are added to the SI loop during a synaptic firing event, and at the maximum value of 3\,\textmu A, roughly 500 fluxons are added. Under these operating conditions, a synaptic firing event with a strong synaptic bias adds 15 times as much current to the SI loop (and therefore the NI loop and NT loop) as a firing event with a weak synaptic bias.
 
In addition to quantifying the current added to the SI loop during a synaptic firing event, we also need to quantify the total storage capacity of the SI loop. To determine this quantity, we use WRSpice to calculate the current in the SI loop as a function of time when $J_{\mathrm{sf}}$ is driven by a fixed current of 10\,\textmu A in addition to the applied $I_{\mathrm{sy}}$. The two traces of Fig.\,\ref{fig:receivers_IsilPerPhoton}(c) show that the saturation current of the SI loop is slightly different for the weak and strong synaptic bias currents. The saturation value of $I_{\mathrm{si}}$ depends on the choice of $L_{\mathrm{sf}}$, and we have chosen $L_{\mathrm{sf}} = 200$\,pH to maximize the saturation value of $I_{\mathrm{si}}$ when $I_{\mathrm{sy}} = 1$\,\textmu A. This maximizes the total number of synaptic firing events the SI loop can store before saturation. It is for this reason that the three-junction circuit of Fig.\,\ref{fig:receivers_schematic}(c) is investigated in this work. Utilizing only one or two junctions results in a decreased storage capacity of the SI loop. In mature designs, it may be advantageous to use a smaller inductor to engineer saturation at a lower level of synaptic activity to introduce an additional nonlinearity to the synapse.

For all WRSpice calculations shown in this work, the value $I_{\mathrm{b}}$ is 9\,\textmu A, leading to current biases across $J_{\mathrm{sf}}$, $J_{\mathrm{jtl}}$, and $J_{\mathrm{si}}$ of 2.2\,\textmu A, 8.1\,\textmu A, and 8.8\,\textmu A, respectively, when the SPD is not firing and $I_{\mathrm{sy}} = 1$\,\textmu A. These numbers are 3.9\,\textmu A, 8.3\,\textmu A, and 8.8\,\textmu A when $I_{\mathrm{sy}} = 3$\,\textmu A. Of these three junctions, only $J_{\mathrm{jtl}}$ is not embedded in a high-inductance loop, making it the most susceptible to noise. This value of $I_{\mathrm{b}}$ has been chosen as a compromise between flux storage capacity of the SI loop and imperviousness to noise. Based on the analysis of Ref.\,\onlinecite{sesc2016}, we calculate the effective temperature, $\tilde{T} = 2\pi k_{\mathrm{B}} T/\Phi_0 I_c$, and inductance parameter, $\lambda = \Phi_0/2\pi LI_c$, where $L$ is the total inductance of the loop. For the junctions under consideration with $I_c = 10$\,\textmu A at 4.2\,K, $\tilde{T} = 0.0176$. The Josephson inductance of the junctions at zero bias is 33\,pH, giving a total loop inductance of 266 pH and an inductance parameter of $\lambda = 0.124$. With these values of $\tilde{T}$ and $\lambda$, the analysis of Ref.\,\onlinecite{sesc2016} informs us that biasing $J_{\mathrm{jtl}}$ with 8.1\,\textmu A is below the switching current of 9\,\textmu A at 4.2\,K. If an application requires further noise reduction, very similar circuit operation can be achieved by applying $I_{\mathrm{b}} = 7$\,\textmu A, provided the range of synaptic biases is shifted to 2\,\textmu A $< I_{\mathrm{sy}} < 4$\,\textmu A. With $I_{\mathrm{b}} = 7$\,\textmu A, the bias across $J_{\mathrm{jtl}}$ is 6.6\,\textmu A when $I_{\mathrm{sy}} = 4$\,\textmu A.

It may be the case that for different applications or during different periods of learning and operation, different amounts of noise are tolerable or even desirable. Changing between $I_{\mathrm{b}} = 7$\,\textmu A and $I_{\mathrm{b}} = 9$\,\textmu A can be done dynamically during operation to modify the stochasticity of the synaptic transducer with no required hardware compensation. Thus, one can utilize or suppress noise at will depending on the context \cite{vrso1996,vora2005,stgo2005}. Further, because synaptic firing events will produce tens to hundreds of fluxons, thermal switching events resulting in the addition of a single fluxon to the SI loop may be inconsequential. The complexity of the effects of noise on the operation of this circuit merit further investigation.

Regarding the two-photon coincidence detector of Fig.\,\ref{fig:receivers_coincidence}, to emulate the physical rebiasing behavior and critical current of the SPDs, JJs with 11\,\textmu A $I_c$ were placed in series with each SPD. The main panel shows simulations of a circuit with $L_{\mathrm{spd}}/r_{\mathrm{spd}} = 1$\,\textmu H$/2\Omega = 500$\,ns, and the inset shows simulations of a circuit with $L_{\mathrm{spd}}/r_{\mathrm{spd}} = 1$\,\textmu H/20\,$\Omega$ = 50\,ns. $I_{\mathrm{b}} = 38$\,\textmu A. For comparison with the circuits of Ref.\,\onlinecite{sh2018c}, the circuit of Fig.\,\ref{fig:receivers_coincidence} has been designed with 40\,\textmu A $I_c$ JJs. Similar performance can be achieved with the 10\,\textmu A junctions used in the circuit of Figs.\,\ref{fig:receivers_simpleCircuit} and \ref{fig:receivers_circuitDiagrams}.

\section{\label{sec:appendix_energy}Energy of a synaptic firing event}
In the receiver circuit explored in this work, each fluxon produced by a synaptic firing event switches three JJs. When combined with the synaptic update circuit or Ref.\,\onlinecite{sh2018c}, a fourth JJ is necessary to act as a buffer between the synaptic update circuit and the synaptic firing circuit. Each switching operation dissipates energy $E_{J} = I_{c} \Phi_{0}$. The total energy dissipated during a synaptic firing event can be minimized by reducing the number of junctions and by reducing their $I_c$, but as we argue in Ref.\,\onlinecite{sh2018d} and \onlinecite{sh2018e}, power dissipation will likely be dominated by light production (unless the light emitters can be made extremely efficient), so reducing the number of junctions or their $I_c$s is not likely to be necessary from an energy perspective. 

The minimum $I_c$ that can be used is determined by noise considerations. For this analysis, we compare the Josephson coupling energy, $I_c\Phi_0/2\pi$, to $k_B T$ with $T = 4.2$\,K \cite{lise1991}. This informs us that the thermal-equivalent junction critical current is 176\,nA. For digital electronics wherein errors are intolerable, most systems employ JJs with $I_c > 40$\,\textmu A. Neuromorphic circuits are resilient to errors and leverage noise \cite{vrso1996,vora2005,stgo2005}. These circuits can therefore operate with lower $I_c$. In the present work, the circuit of Fig.\,\ref{fig:receivers_circuitDiagrams}(a) is designed with all three JJs having $I_c = 10$\,\textmu A ($I_c\Phi_0/2\pi > 50 k_B T$), giving a switching energy of $E_{J} = 21$\,zJ.

The energy of a synaptic firing event is due to rebiasing of the SPD, $E_{\mathrm{spd}} = L_{\mathrm{spd}}I_{\mathrm{spd}}^2/2$, as well as the energy dissipated by the junctions, $E_{\mathrm{JJ}} = 4I_c\Phi_0$. The factor of four comes from the fact that the synaptic firing circuit has four JJs. If we assume the SPD is an out-and-back nanowire integrated with a waveguide \cite{spga2011,pesc2012,feka2015,shbu2017b} with 60\,\textmu m total length, 150\,nm width, and 180\,pH/$\square$, $E_{\mathrm{spd}} = 4$\,aJ. For a synaptic firing event with weak synaptic efficacy (33 fluxons), $E_{\mathrm{JJ}} = 2.7$\,aJ. For a synaptic firing event with strong synaptic efficacy (497 fluxons), $E_{\mathrm{JJ}} = 41$\,aJ. Operation with a weak synaptic weight could be engineered to produce zero fluxons, and a strong synaptic weight one fluxon, thereby reducing the analog transducer to a digital element and achieving the limit of energy efficiency of this part of the neural system. This is likely not necessary as energy consumption is dominated by the production of photons by the transmitter circuit. This contribution to energy consumption is discussed in Ref.\,\onlinecite{sh2018d}.

\section{\label{sec:appendix_ntl}Current induced in neuronal thresholding loop due to current in synaptic integrating loop}
To understand how the synaptic receiver circuits discussed in Secs. \ref{sec:introduction} \textendash \ref{sec:circuitOperation} contribute to the neuron's total integrated current, one must analyze the mutual inductors depicted in Fig. \ref{fig:receivers_circuitDiagrams}(a) and (b). Neuromorphic systems may leverage neurons with only a few synaptic connections or more than a thousand. We consider the cases of $N_{\mathrm{sy}} = $ 10, 100, and 1000. We aim to determine an appropriate choice for $M_{\mathrm{sy}}$, the mutual inductor coupling each SI loop to the NI loop. Because each synaptic connection will contain one of these mutual inductors, we would like it to be as small as possible while still enabling acceptable device performance. By contrast, each neuron will only have a single mutual inductor coupling the NI loop to the NT loop, so its value of mutual inductance can be larger without occupying an intolerable area. Here we consider only symmetrical mutual inductors coupling the SI loop to the NI loop, but we consider a highly asymmetrical mutual inductor coupling the NI loop to the NT loop so that it may provide significant current amplification. We consider a single choice of parameters for the amplifying transformer ($M_{\mathrm{at}}$) coupling the NI loop to the NT loop ($L_{\mathrm{at1}} = 1$\,\textmu H, $L_{\mathrm{at2}} = 100$\,pH, labeled in Fig. \ref{fig:receivers_circuitDiagrams}, and we seek an appropriate value for $M_{\mathrm{sy}}$.

\begin{figure} 
	\centerline{\includegraphics[width=8.6cm]{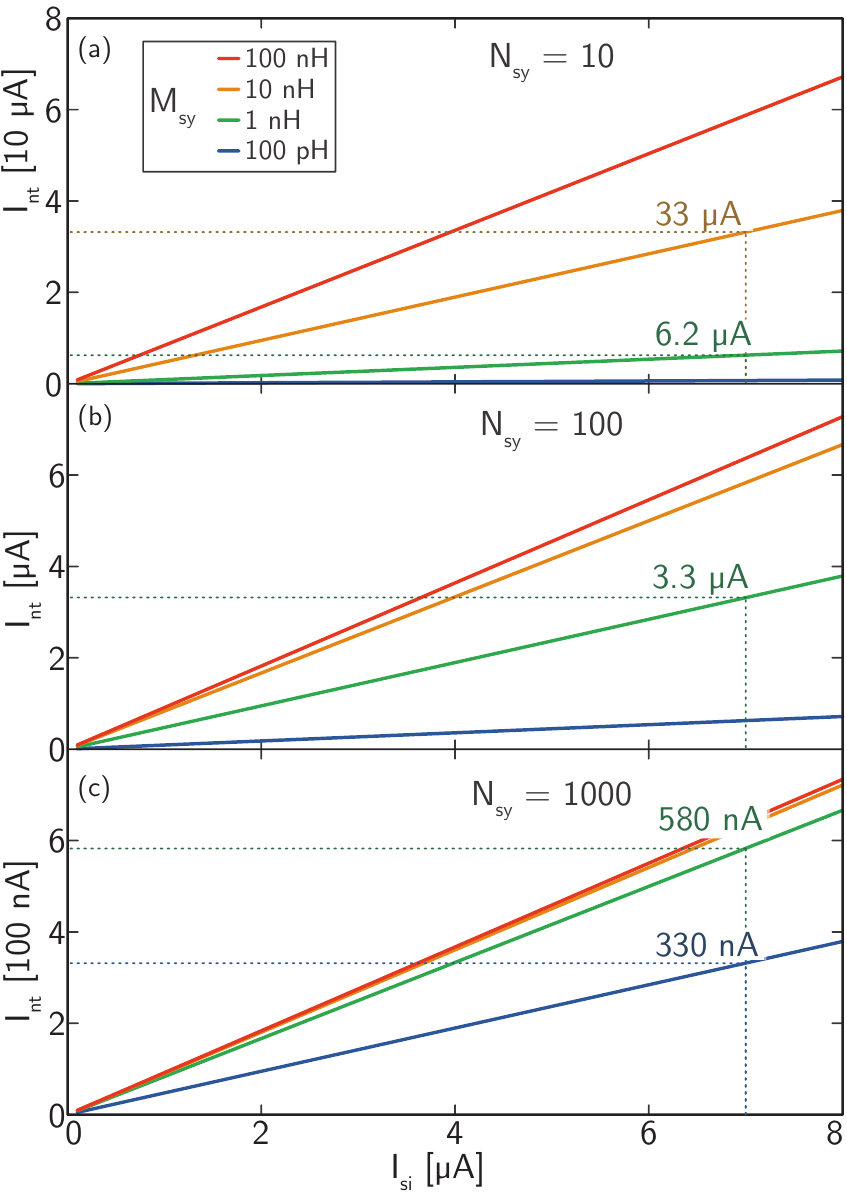}}
	\caption{\label{fig:receivers_IntlVsIsil} Current coupled to the neuronal thresholding loop, $I_{\mathrm{ntl}}$, as a function of the current stored in a synaptic integration loop, $I_{\mathrm{sil}}$, for four values of the mutual inductance, $M_{\mathrm{sy}}$. (a) The number of synaptic connections to the neuronal integration loop, $N_{\mathrm{sy}} = 10$. (b) $N_{\mathrm{sy}} = 100$. (c) $N_{\mathrm{sy}} = 1000$. Synaptic integration loops are observed to saturate just above 7\,\textmu A (Fig. \ref{fig:receivers_IsilPerPhoton}(c)). The currents in the neuronal thresholding loop at this value of current in the synaptic integration loop have been labeled.}
\end{figure}

To determine a functional value for $M_{\mathrm{sy}}$, we calculate the current in the NT loop ($I_{\mathrm{nt}}$) as a function of the current in the SI loop ($I_{\mathrm{si}}$) for values of $M_{\mathrm{sy}}$ spanning three orders of magnitude and for values of $N_{\mathrm{sy}}$ spanning two orders of magnitude. The results of these calculations are shown in Fig. \ref{fig:receivers_IntlVsIsil}.  
In the calculations of Fig. \ref{fig:receivers_IntlVsIsil}, $L_{\mathrm{sil}} = 10$\,\textmu H. The inductors comprising $M_{\mathrm{sy}}$ are equal to $\sqrt{M_{\mathrm{sy}}}$. The neuronal integration loop is assumed to have an additional series inductance of 1 nH. The mutual inductors of the amplifying transformer are $L_{\mathrm{at1}} = 1$\,\textmu H and $L_{\mathrm{at2}} = 100$\,pH. The neuronal thresholding loop is assumed to have an additional series inductance of 10\,pH.

The thresholding element utilized in the NT loop determines the amount of current that must be added to the NT loop to achieve a thresholding event \cite{sh2018d}. Thermal noise will prohibit biasing the device very near $I_c$. As a rough estimate of the current noise in the NT loop, we equate $1/2 L_{\mathrm{ntl}} I_{\mathrm{kT}}^2 = k_{\mathrm{B}} T$ and find $I_{\mathrm{kT}} = 1$\,\textmu A with $L_{\mathrm{ntl}} = 100$\,pH. The choice of threshold is therefore a choice of how much noise the neuron can tolerate. Ion gates in biological neurons are known to have activation near 3 $k_{\mathrm{B}} T$ \cite{daab2001}, and noise applied to neuron activation can be advantageous or even crucial in certain contexts \cite{stgo2005}. The effects of noise on neuromorphic systems are presently incompletely understood. For the purposes of the present work, we assume a current in the NT loop of $I_{\mathrm{th}} = 3-10$\,\textmu A is necessary to achieve a threshold event.

As shown in Sec. \ref{sec:circuitOperation}, the SI loop can store over 7\,\textmu A, even with the minimum synaptic bias. In Fig. \ref{fig:receivers_IntlVsIsil}(a), we see that if we choose $M_{\mathrm{sy}} = $ 1 nH, and there are 10 synaptic connections, when a single SI loop stores 7\,\textmu A, the NT loop will have 6.2\,\textmu A. With $M_{\mathrm{sy}} = $ 10 nH, the NT loop will have 33\,\textmu A. For neurons with a small number of synaptic connections, it is feasible to utilize $M_{\mathrm{sy}} = 10$\,nH. In this case, a single synapse can easily contribute sufficient current to exceed $I_{\mathrm{th}}$.

For the case of $N_{\mathrm{sy}}$ = 100, the smaller size of mutual inductors of $M_{\mathrm{sy}} = 1$\,nH will be advantageous. In this case, a single synapse excited to saturation will only produce 3.3\,\textmu A of current in the NT loop. Thus, a single pre-synaptic neuron exciting the post-synaptic neuron via a single synaptic connection will only be able to drive the neuron to threshold if $I_{\mathrm{th}}\approx 3$\,\textmu A, that is, roughly $3k_{\mathrm{B}} T$. Yet for a neuron with $N_{\mathrm{sy}} = 100$, it is not necessary for a single synaptic connection to be able to drive the neuron to threshold. If as few as 3\% of the synaptic connections are involved in exciting the neuron, $I_{\mathrm{th}}$ can be set at 10\,\textmu A\textemdash comfortably away from thermal noise. Balance between excitation and inhibition in recurrent neural networks leads to synapses with strength of order $1/\sqrt{N_{\mathrm{sy}}}$ \cite{vrso1996,vora2005}, meaning $\sqrt{N_{\mathrm{sy}}}$ excitatory inputs are needed to cross the neuronal firing threshold. Considering the case of a neuron with $N_{\mathrm{sy}} = 1000$, if $\sqrt{N_{\mathrm{sy}}} \approx 32$ synapses are driven to saturation (by a number of synaptic firing events determined by the choice of inductance of the SI loop), 18\,\textmu A would be delivered to NT loop. This current is more than enough to provide a comfortable margin for thresholding above thermal noise \cite{sh2018d}.

The plots in Fig. \ref{fig:receivers_IntlVsIsil} were calculated using the following model. The current present in the NT loop is given by 
\begin{equation}
I_{\mathrm{nt}} = -(M_{\mathrm{at}}/L_{\mathrm{at\_series}})I_{\mathrm{ni}},
\end{equation}
where $M_{\mathrm{at}} = \sqrt{L_{\mathrm{at1}}L_{\mathrm{at1}}}$ and $L_{\mathrm{at\_series}} = L_{\mathrm{at1}}+L_{\mathrm{at1}}$. The current present in the NI loop is given by
\begin{equation}
I_{\mathrm{ni}} = -(M_{\mathrm{sy}}/L_{\mathrm{ni\_tot}})I_{\mathrm{sil}},
\end{equation}
where $L_{\mathrm{ni\_tot}} = L_{\mathrm{ni\_series}}+(N_{\mathrm{_sy}}-1)L_{\mathrm{ni\_lump}}$. $L_{\mathrm{ni\_series}} = M_{\mathrm{sy}}+L_0+L_{\mathrm{at\_lump}}$, where $L_0$ is a parasitic series inductance, modeled as 1\,nH, and $L_{\mathrm{at\_lump}} = L_{\mathrm{at1}}-M_{\mathrm{at}}+M_{\mathrm{at}}(L_{\mathrm{at2}}+L_{\mathrm{at3}}-M_{\mathrm{at}})/(L_{\mathrm{at2}}+L_{\mathrm{at3}})$. Here, $L_{\mathrm{at3}}$ is a parasitic series inductance, modeled as 10\,pH. Finally, $L_{\mathrm{ni\_lump}} = M_{\mathrm{sy}}(L_{\mathrm{si}}+L_{\mathrm{si\_1}}-M_{\mathrm{sy}})/(L_{\mathrm{si}}+L_{\mathrm{si\_1}})+L_{\mathrm{ni\_1}}-M_{\mathrm{sy}}$, where, in this case, $L_{\mathrm{si\_1}} = L_{\mathrm{ni\_1}} = M_{\mathrm{sy}}$.
	
\bibliography{bibliography_modelingSOENs}

\end{document}